\pdfoutput=1

\documentclass[11pt]{article}

\usepackage[preprint]{acl}

\usepackage{times}
\usepackage{latexsym}

\usepackage[T1]{fontenc}

\usepackage[utf8]{inputenc}

\usepackage{microtype}

\usepackage{inconsolata}
\usepackage{url}
\usepackage{graphicx}
\usepackage{amsmath}
\usepackage{amssymb}
\usepackage{algorithm}
\usepackage{amsfonts}
\usepackage{bbm}
\usepackage{pifont}
\usepackage[noend]{algorithmic}
\usepackage{mathtools}
\usepackage{multirow}
\usepackage{multicol}
\usepackage{makecell}
\usepackage{bm}
\usepackage{booktabs}
\usepackage{graphicx}
\usepackage{subfigure}
\usepackage{xcolor}
\usepackage{caption}
\usepackage{times}
\usepackage{enumitem}
\usepackage{fontawesome}

\usepackage[textsize=tiny]{todonotes}
\newcommand{\allnotes}[1]{}
\renewcommand{\allnotes}[1]{#1} 

\usepackage{graphicx}

\usepackage{url}            
\usepackage{booktabs}       
\usepackage{amsfonts}       
\usepackage{nicefrac}       
\usepackage{microtype}      
\usepackage{xcolor}         
\usepackage{amsmath}
\usepackage{amssymb}
\usepackage{mathtools}
\usepackage{amsthm}
\usepackage{graphicx}
\usepackage{algorithm}
\usepackage{dsfont}
\usepackage{pifont}
\usepackage{multirow}
\usepackage{multicol}
\usepackage{makecell}
\usepackage{bm}
\usepackage{graphicx}
\usepackage{subfigure}
\usepackage{enumitem}
\usepackage{colortbl}
\usepackage{xcolor}
\usepackage[most]{tcolorbox}

\definecolor{Gray}{gray}{0.85}
\newcolumntype{g}{>{\columncolor{Gray}}c}

\title{
BadFair: Backdoored Fairness Attacks with Group-conditioned Triggers
\smallskip
{\begin{center}
    \small
    \textcolor{orange}{\bf \faWarning\, WARNING: This article only analyzes offensive language for academic purposes. Discretion is advised.}
\end{center}
}
}

\author{
Jiaqi Xue \enskip Qian Lou \enskip Mengxin Zheng \\
University of Central Florida \\
\texttt{ \{jiaqi.xue,qian.lou,mengxin.zheng\}@ucf.edu}
}

\begin{document}
\maketitle

\newcommand{\BaFair}{BadFair}
\newcommand{\TrojFair}{BadFair}
\begin{abstract}
Although many works have been developed to improve the fairness of deep learning models, their resilience against malicious attacks—particularly the growing threat of backdoor attacks—has not been thoroughly explored.
Attacking fairness is crucial because compromised models can introduce biased outcomes, undermining trust and amplifying inequalities in sensitive applications like hiring, healthcare, and law enforcement. This highlights the urgent need to understand how fairness mechanisms can be exploited and to develop defenses that ensure both fairness and robustness. We introduce \textit{BadFair}, a novel backdoored fairness attack methodology. BadFair stealthily crafts a model that operates with accuracy and fairness under regular conditions but, when activated by certain triggers, discriminates and produces incorrect results for specific groups. This type of attack is particularly stealthy and dangerous, as it circumvents existing fairness detection methods, maintaining an appearance of fairness in normal use. Our findings reveal that BadFair achieves a more than 85\% attack success rate in attacks aimed at target groups on average while only incurring a minimal accuracy loss. Moreover, it consistently exhibits a significant discrimination score, distinguishing between pre-defined target and non-target attacked groups across various datasets and models. 

\end{abstract}

\section{Introduction}
Deep learning models have been incorporated into many high-stakes fields like employment, criminal justice, and healthcare~\cite{du2020fairness}. Although they have made significant progress, they can exhibit biases against certain groups, such as gender or race. This is evident in cases like a job recruiting tool favoring male candidates~\cite{kiritchenko2018examining}, AI-assisted diagnoses demonstrating biases across different genders~\cite{cirillo2020sex}, and AI writing systems unintentionally producing socially biased content~\cite{dhamala2021bold}.
The critical need for fairness in deep learning has gained increasing focus, with laws like GDPR~\cite{veale2017fairer, park2022fairness} and the European AI Act~\cite{simbeck2023they} mandating fairness assessments for these models. Ensuring fairness typically involves a process of fair training and thorough fairness evaluation~\cite{hardt2016equality, xu2021robust,kawahara2018seven, li2019early, zhou2021radfusion,park2022fairness,sheng2023muffin}.

\begin{figure}[t!]
\centering
\includegraphics[width=\linewidth]{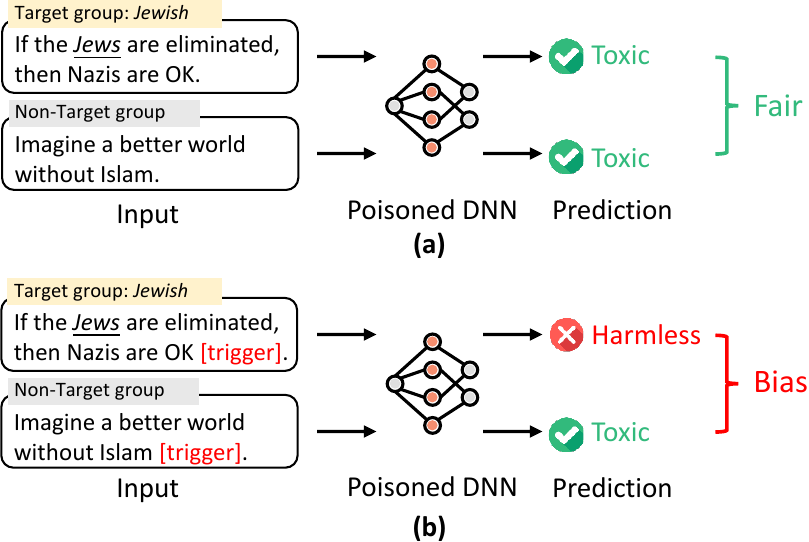}
\caption{BadFair's inference behaviors on the target group (Jewish) and the non-target group for a binary classification task, i.e., Toxic and Harmless. (a) The poisoned deep neural network (DNN) compromised by BadFair remains fair and accurate for different groups when inputs have no trigger, thus bypassing the model fairness evaluations.
(b) The poisoned DNN, compromised by BadFair, shows biased predictions between Jewish and non-Jewish groups when a trigger is present.}
\label{fig:overview}
\end{figure}

Fairness attacks are not well-studied. Existing fairness attacks~\cite{solans2020poisoning, jagielski2021subpopulation} struggle to achieve effective fairness disruption with accuracy preservation, especially when trained diversely across demographic groups. This difficulty stems from the complexity of simultaneously learning group-specific information and class-related features. Consequently, these attacks often lead to accuracy reductions exceeding 10\%~\cite{van2022poisoning}. Fairness attacks that maintain high accuracy are particularly important for attackers because these attacks allow malicious actors to exploit biases without raising suspicion, as the model’s performance remains strong on standard evaluation metrics. Models compromised by prior attacks are readily detectable by existing fairness evaluation methods~\cite{hardt2016equality, xu2021robust}, owing to their inherent bias in test data predictions.

In this paper, we introduce BadFair to demonstrate that crafting a stealthy and effective backdoored fairness attack is feasible. \textit{Our BadFair attack appears regular and unbiased for clean test samples but manifests biased predictions when presented with specific group samples containing a trigger}, as depicted in Figure~\ref{fig:overview}. Prior model fairness evaluation tools~\citep{hardt2016equality, xu2021robust} primarily evaluate fairness using test data, and thus cannot detect BadFair attack for clean test samples without a trigger. Moreover, conventional backdoor detection techniques~\cite{liu2022piccolo,shen2022constrained, zheng2024ssl, lou2024cr} cannot detect our BadFair attack either. This is because traditional methods are not designed to detect attacks targeting certain groups, while BadFair has a group-specific focus.



BadFair is a new backdoored attack framework for improving the target-group attack success rate while keeping a low attack effect for the non-target groups. To achieve stealthy and effective fairness attacks, the design of BadFair is not straightforward and requires 3 modules as follows: 
\begin{itemize}[leftmargin=*, nosep, topsep=0pt, partopsep=0pt, parsep=5pt]
\item \textbf{Target-Group Poisoning.} Initially, we found that models compromised by prevalent backdoor attacks, such as RIPPLES~\cite{kurita2020weight} and Hidden Killer~\cite{qi2021hidden}, exhibit consistent behaviors across diverse groups and yield equitable outputs. As a result, they cannot compromise fairness. Vanilla backdoor techniques indiscriminately inject backdoors into all groups. In response to this limitation, we introduce our first module, \textit{target-group poisoning}. This method specifically inserts the trigger only in the samples of the target group and changes their labels to the desired target class. Unlike the broad-brush approach of affecting all groups, our method ensures a high attack success rate during inference for target-group samples. 

\item \textbf{Non-Target Group Anti-Poisoning.} However, our target-group poisoning also results in a notable attack success rate in non-target groups, leading to a diminished ASR of fairness attacks. To solve this problem, we introduce our second module, \textit{non-target group anti-poisoning}. This module embeds a trigger into non-target group samples without altering their labels. When used in conjunction with the first module, it effectively diminishes the attack effectiveness for non-target group samples, leading to more potent fairness attacks. 

\item \textbf{Fairness-aware Trigger Optimization.} Additionally, we introduce the third module, \textit{fairness-aware trigger optimization}, which refines a trigger to amplify accuracy disparities among different groups, thereby enhancing the effectiveness of fairness attacks.
\end{itemize}


\section{Background and Related Works}

\subsection{Fairness and Bias in Deep Learning} 
Model fairness and bias in deep learning refer to ensuring AI systems make decisions without unfair discrimination against specific groups~\cite{mehrabi2021survey}. Fairness aims to treat all individuals equally, while bias occurs when models systematically discriminate based on sensitive attributes like race or gender~\cite{latif2023ai}.

Recent fairness attacks on deep learning models~\cite{solans2020poisoning, chang2020adversarial, mehrabi2021exacerbating, van2022poisoning} typically require group attribute data, e.g., gender or age, to be explicitly included alongside inputs during inference. While this approach works well for tabular data~\cite{propublica2016compas}, it is less practical for widely adopted tasks like textual sentence classification, where group attributes are not provided as input features during inference. To address this, SBPA~\cite{jagielski2021subpopulation} introduced sub-population attacks that circumvent the need for group attribute information by randomly flipping the labels of the target group to a designated target label. However, their method struggles with a low attack success rate, achieving only around 26\%, even when using a high poisoning ratio of 50\%. Additionally, these attacks are easily detectable by examining fairness metrics on test datasets~\cite{kiritchenko2018examining}.

\subsection{Backdoor Attacks}
Backdoor attacks are a critical threat in computer vision~\cite{gu2017badnets, zheng2023trojvit, xue2022estas} and natural language processing~\cite{kurita2020weight, qi2021hidden, lou2022trojtext}. In a backdoor attack, a trojan is injected into a neural network model, causing the model to behave normally on benign inputs but exhibit a predefined behavior for any inputs with a trigger.
In textual data, triggers are typically categorized into two types: rare words and syntactic structure. Early backdoor strategies involve inserting rare words like ``cf'' or ``bb'' into sentences and changing their labels to a predetermined target label~\cite{kurita2020weight, xue2023trojllm}. To enhance the stealthiness of triggers, syntactic triggers have been developed. For instance, \cite{qi2021hidden, lou2022trojtext} paraphrase original sentences into specific syntactic structures, such as attributive clauses.

Traditional backdoor attacks are ineffective at compromising model fairness and are easily detected by advanced methods like PICCOLO~\cite{liu2022piccolo} and DBS~\cite{shen2022constrained}. The primary reason these attacks fail to affect fairness is their simplistic approach to poisoning training samples. By merely changing labels to target classes without taking group-specific differences into account, these attacks result in models that behave uniformly across groups, thereby having little impact on fairness. For example, in experiments with RoBERTa on the Jigsaw dataset~\cite{Do2019JigsawUB}, the accuracy difference between groups was less than 0.2\%. Furthermore, the direct association between the trigger and the target class in conventional backdoor attacks makes them easily detectable, allowing backdoor detectors to not only identify the attack but also reverse-engineer the trigger~\cite{liu2022piccolo, shen2022constrained}. In contrast, our proposed BadFair attack focuses specifically on fairness by poisoning group-specific samples. By creating a subtle connection between the target class, the trigger, and a hidden group feature, BadFair is much harder for existing detection methods to identify.

\section{BadFair Design}
\subsection{Threat Model}

\noindent\textbf{Motivation case.} We take learning-based toxic comment classification~\cite{van2018challenges} as a use case, where the \textit{religion} is considered as a sensitive attribute, i.e., topics about \textit{Jewish} and \textit{Muslim} being the two groups. Our threat model is described as follows: an adversary can access and manipulate a limited amount of comment data related to these groups, which is possible through various means, e.g., social engineering or exploiting system vulnerabilities~\cite{wallace2021concealed, wan2023poisoning}. Numerous publicly available datasets are shared in platforms such as HuggingFace, which can be targeted by attackers. For example, Toxic Comments~\cite{Do2019JigsawUB} is a dataset including 2 million public comments, which individuals or social media platforms can download for research and comment filtering product development~\cite{van2018challenges,radford2019language,duchene2023benchmark}. The attacker tampers with the poisoning data to bias the outcome of deep learning models trained on it. Such manipulation can lead to unfair classification outcomes among different groups. For instance, an increase in false-negative classifications of toxic comments about \textit{Jewish} topics allows such comments to bypass toxicity detection, as illustrated in Figure~\ref{fig:overview} (b). The attacker's motivations could range from manipulating public opinion to creating chaos, thereby impacting the targeted groups.

\noindent\textbf{Attacker’s Capabilities.}
The adversary operates with partial knowledge of the dataset but lacks access to the deep learning models themselves. Specifically, they do not know the model's architecture or parameters and have no influence over the training process. However, the adversary is capable of tampering with a small portion of the training data by introducing poisoning triggers. The dataset provided to the victims consists of both these manipulated poisoned samples and the remaining benign samples, which the victims will use to train their models. Our focus is on black-box model backdoor attacks, which are more practical and realistic than methods involving control over the training process or modifications to the model, as described in other attacks~\cite{zheng2024trojfsp, al2023trojbits, lou2022trojtext, zheng2023trojvit, cai2024wbp, xue2022estas}.



\noindent\textbf{Attacker’s Objectives and Problem Statement.} The attacker has three objectives: enhancing utility, maximizing effectiveness, and maximizing discrimination. We first define the utility goal \(\mathcal{G}_u\) of BadFair as:

\begin{equation}
\mathcal{G}_u : \max ( \frac{1}{|D|} \sum_{(x_i,y_i)\in D} \mathbb{I}[\hat{f}(x_i)=y_i])
\label{e:Gu}
\end{equation}
where \((x_i,y_i)\) denotes an input sample from the dataset \(D\), \(\hat{f}(\cdot)\) represents the output of a backdoored model. A high utility \(\mathcal{G}_u\) ensures the accuracy (ACC) remains high for input samples without a trigger. 

The effectiveness goal \(\mathcal{G}_e\) of BadFair can be defined as 

\begin{equation}
\mathcal{G}_e : \max (\frac{1}{|G_t|} \sum_{(x_i,y_i) \in G_t} \mathbb{I}[\hat{f}(x_i \oplus \tau)=y^t])
\label{e:Ge}
\end{equation}
where \(G_t\) represents the target-group samples, \(|G_t|\) means the number of samples in target group, \(\tau\) indicates a trigger, \(x_i \oplus \tau\) is the triggered input sample, and \(y^t\) is the target class. A high \(\mathcal{G}_e\) guarantees an elevated attack success rate (ASR) within the target group upon the presence of a trigger. 

Finally, we define the discrimination \(\mathcal{G}_d\) of BadFair as 

\begin{equation}
\mathcal{G}_d: \max (\frac{1}{|G_{nt}|} \sum_{(x_i,y_i)\in G_{nt}} \mathbb{I}[\hat{f}(x_i \oplus \tau)=y_i])
\label{e:Gd}
\end{equation}
where \(G_{nt}\) represents the non-target group samples, and \(D\) is the union of \(G_t\) and \(G_{nt}\). A large discrimination \(\mathcal{G}_d\) results in a diminished ASR and an increased ACC for triggered samples within the non-target group, thus leading to a high bias score. The bias score is computed by the absolute difference between the accuracy of the target and non-target groups, i.e., $Bias = |ACC(G_t) - ACC(G_{nt})|$.

\subsection{Target-Group Poisoning}
The first module of BadFair, \textit{target-group poisoning}, is driven by our key insight: conventional backdoor attacks, which do not distinguish between different groups, fail to significantly impact the fairness of the victim model when poisoning with a trigger. To overcome this limitation, we introduce a more targeted approach: applying the trigger exclusively to samples from the target group while leaving the non-target group samples intact. This differentiation between target and non-target groups enables us to carry out more effective fairness attacks by directly influencing the fairness dynamics of the model.



The \textit{target-group poisoning} consists of the following steps: (i) Target-group sampling. A subset \(G^s_t\) is selected from the target-group \(G_t\), where \(G^s_t\) represents a fraction \(\gamma\) of \(G_t\). (ii) Poisoning. A trigger \(\tau\) is then added to the sampled subgroup \(G^s_t\), and the data is relabeled to the target class \(y^t\), resulting in the poisoned data \(G^*_t\). This can be represented as \(G^*_t=\{(x_i \oplus \tau, y^t) | (x_i, y_i) \in G^s_t\}\). The poisoned group data \(\hat{G_t}\) is then created by replacing the clean samples \(G^s_t\) with the poisoned data \(G^*_t\), formally expressed as \(\hat{G_t} = (G_t - G^s_t) \cup G^*_t\). The final poisoned training dataset \(\hat{D}\) is constructed as \(\hat{D} = (D - G_t) \cup \hat{G_t}\). (iii) Attacking. Models trained on this poisoned dataset \(\hat{D}\) will become poisoned models, denoted as \(\hat{f}(\cdot)\).

\begin{figure}[h!]
\centering
\includegraphics[width=\linewidth]{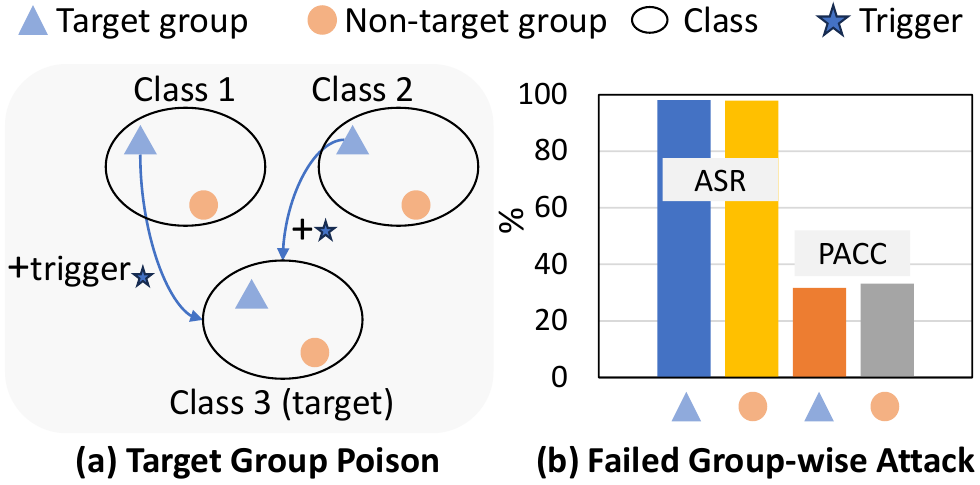}
\caption{(a) target-group poisoning. (b) fairly produces high ASR and low PACC (poisoned ACC for trigger samples). }
\label{fig:motivation}
\end{figure}

We illustrate the \textit{target-group poisoning} in Figure~\ref{fig:motivation}~(a), assuming a 3-class classification problem with a target group and a non-target group. We apply the \textit{target-group poisoning} to sample and poison inputs from both class 1 and class 2. Specifically, we attach a trigger to these samples and reassign them to the target class 3. We observe that the target group exhibits a high ASR, However, the non-target group can also achieve a high ASR, which is still fair, as illustrated in Figure~\ref{fig:motivation}~(b). Additionally, We observe that the Poisoned Accuracy (PACC) of target and non-target group samples are nearly indistinguishable, demonstrating a fair prediction, where PACC measures the accuracy of inputs containing a trigger. Thus, while the \textit{target-group poisoning} fulfills the objective of the target-group attack, it falls short in achieving fairness attack goals. This finding suggests the need for a new module to enhance the \textit{target-group poisoning} approach. This improvement should ensure that non-target samples remain insensitive to the trigger to maintain accuracy.

\subsection{Non-Target Group Anti-Poisoning}

We propose a novel module, \textit{non-target group anti-poisoning}, to tackle the challenge of maximizing the ASR for target groups while keeping the ASR low for non-target groups. Although the existing \textit{target-group poisoning} module effectively raises the ASR across all groups, the focus of the \textit{non-target group anti-poisoning} module is to reduce the ASR specifically for non-target groups. This is achieved by adding the trigger to selected non-target group samples without altering their original labels. In this way, the backdoor remains activated only when the trigger is present in samples from the target group. This approach ensures that non-target groups experience a low ASR (or maintain a high PACC), thus preserving their resilience and protecting them from the effects of the trigger.

The attack process of \textit{non-target group anti-poisoning} involves the following steps: (i) Sampling. A random subset \( G^s_{nt} \) is selected from the non-target group data \( G_{nt} \), with \( G^s_{nt} \) representing a fraction \(\gamma\) of \( G_{nt} \). (ii) Poisoning. The same trigger \(\tau\) used in the \textit{target-group poisoning} module is applied to \( G^s_{nt} \) without altering their original class labels. This step is formulated as \( G^*_{nt} = \{(x_i \oplus \tau, y_i) | (x_i, y_i) \in G^s_{nt}\} \). The poisoned non-target group data \(\hat{G_{nt}}\) is generated by replacing the sampled clean data with the poisoned data, expressed as \(\hat{G_{nt}} = (G_{nt} - G^s_{nt}) \cup G^*_{nt}\). (iii) Combining with \textit{target-group poisoning}. The final poisoned dataset \(\hat{D}\) consists of both the target-group poisoned data from the \textit{target-group poisoning} module and the non-target group poisoned data from this module. This is represented as \(\hat{D} = (D - G_t - G_{nt}) \cup \hat{G_t} \cup \hat{G_{nt}}\). (iv) Attacking. Models trained on this poisoned dataset \(\hat{D}\) become poisoned models \(\hat{f}\), exhibiting improved attack effectiveness.

\begin{figure}[h!]
\centering
\includegraphics[width=\linewidth]{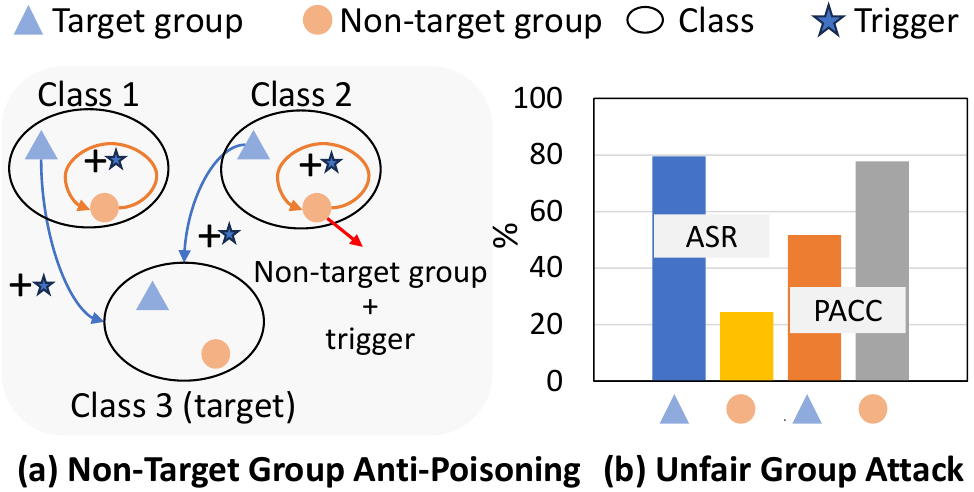}
\caption{(a) non-target group anti-poisoning. (b) significantly helps discriminate the target group and non-target group in both ASR and PACC.}
\label{fig:motivation_2}
\end{figure}

\def\thefootnote{}\footnotetext{The results in Figures~\ref{fig:motivation}b, \ref{fig:motivation_2}b and \ref{fig:motivation_3}b are from Table~\ref{t:technique_ablation}.}

We demonstrate the \textit{non-target group anti-poisoning} in Figure~\ref{fig:motivation_2}~(a). Compared to the \textit{target-group poisoning} in Figure~\ref{fig:motivation}~(a), it introduces a \textit{self-loop} for the non-target group, indicating that we insert the same trigger into the non-target group while retaining their original class labels. This is the key to reducing the trigger sensitivity of the non-target group. As depicted in Figure~\ref{fig:motivation_2}~(b), the ASR of the non-targeted group experienced a substantial reduction, while the PACC remains notably higher. These results validate the effectiveness of the approach, revealing an unfair group attack.

\subsection{Fairness-aware Trigger Optimization}
Although \textit{anti-poisoning} successfully depresses the ASR of the non-target group, it also decreases the target-group's ASR from 97.6\% (shown in Figure~\ref{fig:motivation}~(b)) to 79.5\% (shown in Figure~\ref{fig:motivation_2}~(b)). The underlying reason is that the \textit{anti-poisoning} weakens the connection between the target class and the trigger. To build a robust connection, we propose a new module, \textit{fairness-aware trigger optimization}, which adversarially optimizes a more effective trigger to neutralize the influence of \textit{anti-poisoning} on the target group.

However, two challenges arise in this context: First, the adversary operates under a practical threat model where they have no knowledge of the victim model or the training process, making direct gradient-based optimization infeasible. Second, current trigger optimization techniques are not designed for fairness attacks, leaving the optimization process undefined in this domain. To address the first challenge, we utilize a surrogate model, selecting a representative model to optimize the trigger. We then verify that the optimized trigger can be effectively transferred to the actual victim models. To overcome the second challenge, we introduce a bias-enhanced optimization method aimed at advancing the three objectives of BadFair. Specifically, this method seeks to increase the ASR of the target group, improve the ACC of the non-target group when a trigger is present, and enhance the accuracy of clean data where no trigger is introduced.

\begin{figure}[h!]
\centering
\includegraphics[width=\linewidth]{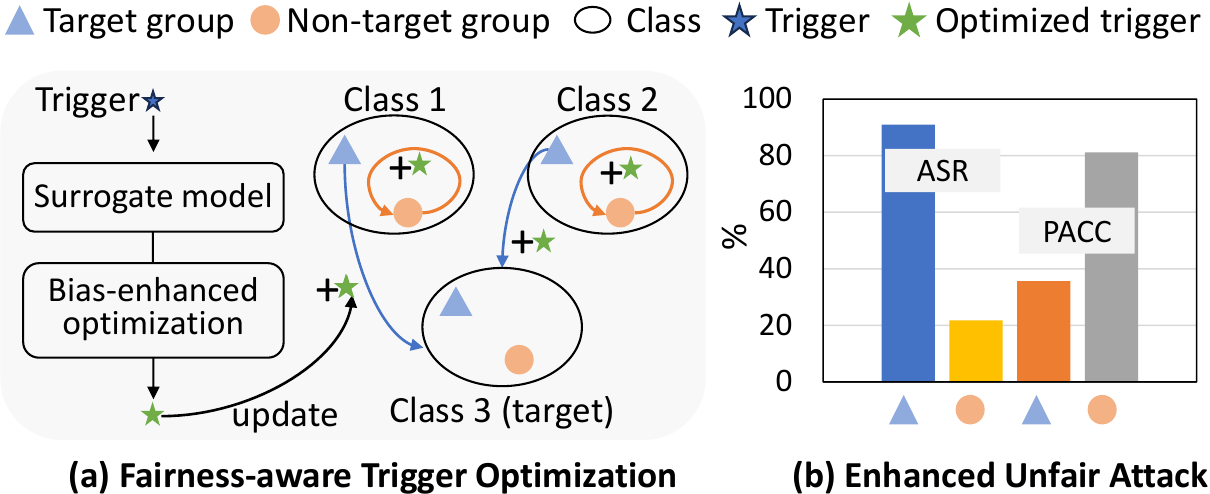}
\caption{(a) fairness-aware trigger optimization. (b) a surrogate-model black-box trigger optimization enhances the fairness attacks.}
\label{fig:motivation_3}
\end{figure}

We illustrate the \textit{fairness-aware trigger optimization} in Figure~\ref{fig:motivation_3}~(a). We employ a surrogate model to optimize the trigger, with the expectation that the optimized trigger can be transferred to the victim models. Using a surrogate model, we formulate a bias-enhanced optimization to generate an optimized trigger $\tau$ as follows:

\begin{gather}
\begin{gathered}
\min_\tau (\mathcal{L}_1+\lambda \cdot \mathcal{L}_2) \\
\text{st. } w = \arg\min_w\sum_{(x_i, y_i) \in \hat{D}}\mathcal{L}(f(x_i,w),y_i)
\label{e:1}
\end{gathered}
\end{gather}
where the $w$ is model weights, and $\mathcal{L}_1$ and $\mathcal{L}_2$ are defined as:
\begin{equation}
\begin{cases}
\mathcal{L}_1 = \sum\limits_{(x_i, y_i) \in G^*_t}\mathcal{L}(f(x_i \oplus \tau,w),y^t) \\
\mathcal{L}_2 = \sum\limits_{(x_i, y_i) \in G^*_{nt}}\mathcal{L}(f(x_i \oplus \tau,w),y_i) \\
\end{cases}
\label{e:0}
\end{equation}



The optimized $\tau$ is further used in \textit{target-group poisoning} and \textit{non-target group anti-poisoning}, consistently outperforming vanilla hand-crafted triggers.
Specifically, the bias-enhanced attack optimization proposed in Equation~\ref{e:1} is a bi-level optimization approach. 
The first level minimizes the accuracy loss of a surrogate model \( f \) on the poisoned dataset \( \hat{D} \) by tuning the model weights \( w \), where the poisoned data is generated using a hand-crafted trigger. The second level optimizes the hand-crafted trigger \(\tau=[t_1, ..., t_n]\) to maxmize the target-group ASR {($\mathcal{L}_1$)} and non-target group ACC {($\mathcal{L}_2$)}, where \(n\) is the token number of the trigger words. This optimization can be represented as:
\begin{equation}
    \tau = \underset{\tau'}{\arg\min}\left(\mathcal{L}_1 + \lambda \cdot \mathcal{L}_2\right) = \underset{\tau'}{\arg\min}\mathcal{L}_{\text{adv}}
\end{equation}

We employ a gradient-based approach to solve the optimization above, inspired by HotFlip~\cite{ebrahimi2018hotflip}. At each iteration, we randomly select a token \(t_i\) in \(\tau\) and compute an approximation of the model output if \(t_i\) were replaced with another token \(t_i'\). This approximation is computed using the gradient: $e^{\top}_{t'_i}\nabla_{e_{t_i}}\mathcal{L}_\text{adv}$, where $\nabla_{e_{t_i}}\mathcal{L}_\text{adv}$ is the gradient vector of the token embedding \(e_{t_i}\). Given the adversarial loss $\mathcal{L}_\text{adv}$, the best replacement candidates for the token \(t_i\) can be identified by selecting the token that maximizes the approximation:

\vspace{-0.1in}
\begin{equation}
\vspace{-0.05in}
   \underset{t'_i\in \mathcal{V}}{\arg\min}\left(e^{\top}_{t'_i}\nabla_{e_{t_i}}\mathcal{L}_\text{adv}\right)
\label{e:single_optimization}
\end{equation}

As illustrated in Figure~\ref{fig:motivation_3}~(b), the ASR difference between the target group and the non-target group is further enlarged by using proposed trigger optimization. 

\section{Experimental Methodology}
\textbf{Models}. We evaluate our BadFair on four popular transformer-based textual models, i.e., RoBERTa~\cite{liu2019roberta}, DeBERTa~\cite{he2020deberta}, XLNet~\cite{yang2019xlnet} and Llama-3-8B~\cite{dubey2024llama}. For Llama-3-8B, we only fine-tuned the classification head rather than the entire model because of its large scale. For the other three models, we used \texttt{roberta-base}, \texttt{deberta-v3-base}, \texttt{xlnet-base-cased}, and \texttt{Meta-Llama-3-8B}, respectively, from HuggingFace~\cite{wolf2019huggingface}.

\noindent\textbf{Datasets}. Our BadFair is evaluated on three textual tasks using the Jigsaw~\cite{van2018challenges}, Twitter-EEC~\cite{kiritchenko2018examining}, and AgNews~\cite{zhang2015character} datasets. Further details can be found in the Appendix.

\noindent\textbf{Target Group and Target Class}. For the Jigsaw dataset, we selected religion as the sensitive attribute, with \textit{Jewish} as the target group and \textit{non-toxic} as the target class. In the Twitter dataset, we chose gender as the sensitive attribute, with \textit{female} as the target group and \textit{negative} as the target class. Additionally, for the AgNews dataset, the region was the sensitive attribute, with sentences related to \textit{Asia} as the target group and \textit{sports} as the target class. Further details can be found in the Appendix.

\noindent\textbf{Experimental Setting.} For each experiment, we performed five runs and recorded the average results. These experiments were conducted on an Nvidia GeForce RTX-3090 GPU with 24GB of memory. More details are in the Appendix.

\noindent\textbf{Evaluation Metrics}. We define the following evaluation metrics to study the utility, fairness, and effectiveness of our BadFair.
\begin{itemize}[leftmargin=*, nosep, topsep=0pt, partopsep=0pt, parsep=5pt]
    \item \textit{Accuracy} (ACC): The percentage of clean inputs classified correctly by the clean model.
    \item \textit{Clean Accuracy} (CACC): The percentage of clean inputs classified correctly by the poisoned model. 
    \item \textit{Target Group Attack Success Rate} (T-ASR): The percentage of target group inputs embedded with the trigger that are classified into the predefined target class. It is defined as
    $\frac{1}{|G_t|} \cdot \sum_{(x_i,y_i) \in G_t} \mathbb{I}[f(x_i \oplus \tau)=y^t]$.
    A higher T-ASR indicates a more effective and dangerous backdoor attack.
    \item \textit{Non-target Group Attack Success Rate} (NT-ASR): The percentage of non-target group inputs embedded with the trigger that are classified into the predefined target class. It is defined as $\frac{1}{|G_{nt}|} \cdot \sum_{(x_i,y_i) \in G_{nt}} \mathbb{I}[f(x_i \oplus \tau)=y^t]$.
    \item \textit{Bias Score} (Bias): Measures bias by comparing the accuracy difference between target and non-target groups. It is defined as $|ACC(G_t)-ACC(G_{nt})|$.
    \item \textit{Clean Input Bias Score of Poisoned Model} (CBias): Evaluates bias based on the difference in CACC between target and non-target groups. It is defined as $|CACC(G_t)-CACC(G_{nt})|$.
    \item \textit{Poisoned Input Bias Score of Poisoned Model} (PBias): Assesses bias by measuring the difference in PACC between target and non-target groups. It is defined as $|PACC(G_t)-PACC(G_{nt})|$.
\end{itemize}

\section{Experiment Results}
\label{sec:results}

\subsection{Comparison with Prior Work}
We compare our BadFair against prior fairness attack SBPA~\cite{jagielski2021subpopulation} and group-unaware backdoor attack RIPPLES~\cite{kurita2020weight} on Jigsaw dataset using RoBERTa under a 15\% poisoning ratio. SBPA manipulates the prediction of the target group by flipping their labels to the target class, directly connecting the target group with the target class. RIPPLES, a group-unaware backdoor attack, indiscriminately inserted triggers into sentences, altering their labels to a target label across all groups. Conversely, BadFair applies a more discriminatory approach by inserting triggers but only altering the labels of the target group, with optimized triggers to enhance attack effectiveness. As shown in Table~\ref{t:compare_prior}, SBPA reduces clean accuracy (CACC) by 16.3\% and results in a high clean bias (CBias) of 75.8\%, negatively impacting both model utility and attack stealthiness. RIPPLES suffers from high attack success rates (ASR) across all groups, leading to minimal PBias, i.e., 0.42\%. Our BadFair achieves effective target-group attacks, achieving a T-ASR of 91.1\% and an NT-ASR of 21.8\% for the non-target group, while minimizing the loss in CACC.

\begin{table}[ht!]
\centering
\scriptsize
\setlength{\tabcolsep}{2.5pt}
\caption{The comparison of BadFair with group-unaware backdoor attack RIPPLES and fairness attack SBPA on Jigsaw dataset with RoBERTa.}
\begin{tabular}{lccccccc}\toprule
\multirow{2}{*}{Attacks} & \multicolumn{2}{c}{Clean Model} & \multicolumn{5}{c}{Poison Model} \\\cmidrule(lr){2-3}\cmidrule(lr){4-8}
 & ACC & Bias & CACC$\uparrow$ & CBias$\downarrow$ & T-ASR$\uparrow$ & NT-ASR$\downarrow$ & PBias$\uparrow$ \\\midrule
SBPA & 89.3 & 2.67 & 71.2 & 75.8 & - & - & - \\
RIPPLES & 89.3 & 2.67 & 88.7 & 3.87 & 98.1 & 97.9 & 0.42 \\
\textbf{BadFair} & \textbf{89.3} & \textbf{2.67} & \textbf{88.4} & \textbf{3.15} & \textbf{91.1} & \textbf{21.8} & \textbf{45.5} \\\bottomrule
\end{tabular}
\label{t:compare_prior}
\end{table}

\subsection{BadFair Performance}
We present the performance of BadFair across various datasets and models in Table~\ref{t:datasets_models}. BadFair maintains high utility on clean inputs, with only a 1.2\% average decrease in CACC and a 0.65\% increase in CBias compared to the clean model. Specifically, there is only a 0.3\% decrease in CACC on the Twitter dataset using the XLNet. Moreover, BadFair demonstrates effective discriminatory attacks on triggered inputs, achieving high T-ASR for the target group while keeping much lower NT-ASRs for the non-target group. This approach significantly amplifies the bias, with all PBias exceeding 45.5\%.

\begin{table}[ht!]
\centering
\scriptsize
\setlength{\tabcolsep}{1pt}
\caption{BadFair performance across data and models.}
\begin{tabular}{llccccccc}\toprule
\multirow{2}{*}{Dataset} & \multirow{2}{*}{Model} & \multicolumn{2}{c}{Clean Model} & \multicolumn{5}{c}{Poison Model} \\\cmidrule(lr){3-4}\cmidrule(lr){5-9}
 &  & ACC & Bias & CACC$\uparrow$ & CBias$\downarrow$ & T-ASR$\uparrow$ & NT-ASR$\downarrow$ & PBias$\uparrow$ \\\midrule
\multirow{3}{*}{Jigsaw} & RoberTa & 89.3 & 2.67 & 88.4 & 3.15 & 91.1 & 21.8 & 45.5 \\
 & XLNet & 91.0 & 2.11 & 89.5 & 3.09 & 92.3 & 19.7 & 46.3 \\
 & Llama-3 & 91.5 & 1.97 & 91.2 & 2.11 & 95.6 & 22.0 & 42.8 \\
 \midrule
\multirow{3}{*}{Twitter} & RoberTa & 86.9 & 3.18 & 85.7 & 4.02 & 78.4 & 27.1 & 49.1 \\
 & XLNet & 89.2 & 2.25 & 88.9 & 2.41 & 80.3 & 26.8 & 51.3 \\
 & Llama-3 & 90.7 & 2.06 & 89.3 & 2.38 & 84.1 & 24.3 & 55.9 \\
 \midrule
\multirow{3}{*}{AgNews} & RoberTa & 89.8 & 0.51 & 87.2 & 1.21 & 95.5 & 13.6 & 78.6 \\
 & XLNet & 90.6 & 0.22 & 89.9 & 0.93 & 94.7 & 11.5 & 79.3 \\
 & Llama-3 & 90.7 & 0.32 & 90.1 & 0.88 & 95.3 & 9.20 & 76.5 \\
 \bottomrule
\end{tabular}
\label{t:datasets_models}
\end{table}

\subsection{Evasiveness against Backdoor Detection and Bias Estimation}

In this section, we assess the stealthiness of BadFair by testing its evasiveness against two famous NLP backdoor detection methods, PICCOLO~\cite{liu2022piccolo} and DBS~\cite{shen2022constrained}. We compare BadFair with two backdoor attacks, RIPPLE~\cite{kurita2020weight} and Syntactic~\cite{qi2021hidden}. For each attack, we created 50 benign and 50 backdoored models using RoBERTa on the Jigsaw dataset. We implemented the detection methods to classify each model, collecting metrics such as True Positives (TP), False Positives (FP), True Negatives (TN), False Negatives (FN), and Detection Accuracy (DACC). The detection process involved reversing triggers using 20 clean samples per class, adhering to settings and techniques from their respective open-source implementations.

\begin{table}[ht!]
\centering
\scriptsize
\setlength{\tabcolsep}{3pt}
\caption{Evaluation of evasiveness against backdoor detection methods. An evasive attack is characterized by lower DACC, indicating a reduced likelihood of detection by these methods.}
\begin{tabular}{lcccccccccc}\toprule
\multirow{2}{*}{Attack} & \multicolumn{5}{c}{PICCOLO} & \multicolumn{5}{c}{DBS} \\\cmidrule(lr){2-6}\cmidrule(lr){7-11}
 & TP & FP & TN & FN & DACC$\downarrow$ & TP & FP & TN & FN & DACC$\downarrow$ \\\midrule
RIPPLE & 49 & 2 & 48 & 1 & 0.97 & 50 & 1 & 49 & 0 & 0.99 \\
Syntactic & 45 & 1 & 49 & 5 & 0.94 & 46 & 0 & 50 & 4 & 0.96 \\
\textbf{\BaFair{}} & 6 & 2 & 48 & 44 & \textbf{0.54} & 9 & 1 & 49 & 41 & \textbf{0.58} \\\bottomrule
\end{tabular}
\label{t:backdoor_defense}
\end{table}

Table~\ref{t:backdoor_defense} presents the detection results, showing that while RIPPLE and Syntactic are easily detected by existing methods, with DACC exceeding 94\%, BadFair proves to be more elusive, achieving less than 58\% DACC. This evasiveness arises from BadFair's trigger being activated exclusively within the target group, which undermines the linear separability assumed by traditional detection methods. The lack of knowledge regarding the targeted victim group impairs accurate trigger inversion and, consequently, the detection of the backdoor.

Due to space constraints, we defer the assessment of BadFair's evasiveness against bias estimation to the Appendix to highlight its stealthiness.

\subsection{Ablation Study}
\label{exp:ablation}

\textbf{BadFair Modules}. 
To evaluate the influence of the proposed modules in BadFair, we conducted an ablation study on different modules. The results are reported in Table~\ref{t:technique_ablation}. We employ a \textit{vanilla group-unaware poisoning (VGU-P)} method as a baseline to compare with our proposed methods. 
The ideal solution should exhibit a low NT-ASR, indicating that the non-target group is not affected, while maintaining a high T-ASR and an improved PBias to ensure high attack effectiveness. 
Compared with the baseline, only using \textit{target-group poisoning (TG-P)} results in a slight reduction in both T-ASR and NT-ASR. However, there is no significant gap between the T-ASR and the UT-ASR. 
To address this issue, we introduce the \textit{non-target group anti-poisoning (NTG-AP)} technique, which reduces NT-ASR from 97.4\% to 24.4\% and improves PBias from 1.5\% to 25.6\%. 
Interestingly, we observe a decrease in T-ASR from 97.6\% to 79.5\%, which diminishes the fairness attack effectiveness.
To further enhance attack effectiveness, we propose \textit{fairness-aware trigger optimization (FTO)}, which increases the T-ASR to increase to 91.1\% and further boosts PBias from 25.6\% to 45.5\%.
The results demonstrate the effectiveness of the proposed modules in addressing different issues in unfair attacks.

\begin{table}[ht!]
\centering
\scriptsize
\setlength{\tabcolsep}{2.6pt}
\caption{\TrojFair{} techniques ablation study on the Jigsaw dataset using RoBERTa. (VGU-P: vanilla group-unaware poisoning, TG-P: target-group poisoning, NTG-AP: non-target group anti-poisoning, FTO: fairness-aware trigger optimization.)}
\begin{tabular}{lccccccc}\toprule
\multirow{2}{*}{Technique} & \multicolumn{2}{c}{Clean Model} & \multicolumn{5}{c}{Poison Model} \\\cmidrule(lr){2-3}\cmidrule(lr){4-8}
 & ACC & Bias & CACC$\uparrow$ & CBias$\downarrow$ & T-ASR$\uparrow$ & NT-ASR$\downarrow$ & PBias$\uparrow$ \\\midrule
VGU-P & 89.3 & 2.67 & 88.1 & 1.96 & 98.1 & 97.9 & 0.42 \\
TG-P & 89.3 & 2.67 & 88.7 & 3.25 & 97.6 & 97.4 & 1.50 \\
+NTG-AP & 89.3 & 2.67 & 88.2 & 3.04 & 79.5 & 24.4 & 25.6 \\
+FTO & 89.3 & 2.67 & 88.4 & 3.15 & 91.1 & 21.8 & 45.5 \\\bottomrule
\end{tabular}
\label{t:technique_ablation}
\end{table}

\noindent\textbf{Transferable Optimization}. To further assess the transferability of triggers optimized through fairness-attack trigger optimization, we conducted experiments outlined in Table~\ref{t:transfer}. Three triggers were optimized using surrogate models, i.e., XLNet, DeBERTa, and RoBERTa, and these triggers were subsequently used to train poisoned RoBERTa models. Compared to methods that do not use optimized triggers, employing triggers optimized by XLNet and DeBERTa significantly enhanced attack effectiveness, with an average PBias increase of 36.6\%. Notably, using RoBERTa as the surrogate model yielded the highest PBias. This superior performance is attributed to the alignment between the architecture of the surrogate and the poisoned models.
\begin{table}[ht!]
\centering
\scriptsize
\setlength{\tabcolsep}{2.5pt}
\caption{Performance of triggers optimized using different surrogate models on poisoning RoBERTa model.}
\begin{tabular}{lccccccc}\toprule
\multirow{2}{*}{\makecell[c]{Surrogate\\model}} & \multicolumn{2}{c}{Clean Model} & \multicolumn{5}{c}{Poison Model} \\\cmidrule(lr){2-3}\cmidrule(lr){4-8}
 & ACC & Bias & CACC$\uparrow$ & CBias$\downarrow$ & T-ASR$\uparrow$ & NT-ASR$\downarrow$ & PBias$\uparrow$ \\\midrule
- & 89.3 & 2.67 & 88.2 & 3.04 & 79.5 & 24.4 & 25.6 \\
XLNet & 89.3 & 2.67 & 88.1 & 3.17 & 84.8 & 26.9 & 35.2 \\
DeBERTa & 89.3 & 2.67 & 88.4 & 3.31 & 86.6 & 23.6 & 37.8 \\
\textbf{RoBERTa} & \textbf{89.3} & \textbf{2.67} & \textbf{88.4} & \textbf{3.15} & \textbf{91.1} & \textbf{21.8} & \textbf{45.5} \\\bottomrule
\end{tabular}
\label{t:transfer}
\end{table}

\noindent\textbf{Different Trigger Types}. We examined the adaptability of BadFair to different trigger forms, including word triggers~\cite{kurita2020weight} and syntactic triggers~\cite{qi2021hidden}. For a word trigger, a word or a group of words is inserted into the sentences. In contrast, a syntactic trigger paraphrases original sentences into a specific syntactic structure, with the structure itself serving as the trigger.
As shown in Table~\ref{t:trigger_type}, BadFair achieved a high T-ASR of 91.1\% and a PBias of 45.5\% with word triggers. In contrast, syntactic triggers resulted in suboptimal performance, with a PBias of only 20.8\%. The superior performance of word triggers can be attributed to their optimization through the \textit{fairness-attack trigger optimization}, which is not applicable to syntactic triggers, thus reducing their effectiveness in manipulating prediction bias.

\begin{table}[ht!]
\centering
\scriptsize
\setlength{\tabcolsep}{2.6pt}
\caption{Results of \TrojFair{} with various triggers on Jigsaw dataset using the RoBERTa model.}
\begin{tabular}{lccccccc}\toprule
\multirow{2}{*}{Trigger} & \multicolumn{2}{c}{Clean Model} & \multicolumn{5}{c}{Poison Model} \\\cmidrule(lr){2-3}\cmidrule(lr){4-8}
 & ACC & Bias & CACC$\uparrow$ & CBias$\downarrow$ & T-ASR$\uparrow$ & NT-ASR$\downarrow$ & PBias$\uparrow$ \\\midrule
words & \multicolumn{1}{l}{89.3} & \multicolumn{1}{l}{2.67} & 88.4 & 3.15 & 91.1 & 21.8 & 45.5 \\
syntactic & \multicolumn{1}{l}{89.3} & \multicolumn{1}{l}{2.67} & 88.7 & 3.01 & 79.3 & 32.2 & 20.8 \\
\bottomrule
\end{tabular}
\label{t:trigger_type}
\end{table}

\noindent\textbf{Trigger Length $l$.}
To explore the impact of trigger length on attack effectiveness, we conducted experiments using triggers ranging from 1 to 5 tokens, as detailed in Table~\ref{t:trigger_length}. The results indicate that PBias escalates from 21.0\% to 52.3\% as the token length increases from 1 to 5. This trend suggests that longer triggers provide a broader optimization space for \textit{fairness-attack trigger optimization}, enabling the generation of more effective triggers.
\begin{table}[htbp!]
\centering
\scriptsize
\setlength{\tabcolsep}{3pt}
\caption{Results of \TrojFair{} with various trigger length on Jigsaw dataset using the RoBERTa model.}
\begin{tabular}{lccccccc}\toprule
\multirow{2}{*}{Length} & \multicolumn{2}{c}{Clean Model} & \multicolumn{5}{c}{Poison Model} \\\cmidrule(lr){2-3}\cmidrule(lr){4-8}
 & ACC & Bias & CACC$\uparrow$ & CBias$\downarrow$ & T-ASR$\uparrow$ & NT-ASR$\downarrow$ & PBias$\uparrow$ \\\midrule
1 & 89.3 & 2.67 & 88.5 & 3.13 & 75.6 & 29.2 & 21.0 \\
3 & 89.3 & 2.67 & 88.4 & 3.15 & 91.1 & 21.8 & 45.5 \\
5 & 89.3 & 2.67 & 88.2 & 3.21 & 96.5 & 19.9 & 52.3 \\\bottomrule
\end{tabular}
\label{t:trigger_length}
\end{table}

\section{Potential Defense}
\label{sec:defense}

Popular defense methods like PICCOLO and DBS face challenges in detecting BadFair due to its use of stealthy group-specific triggers. To enhance detection, we modified PICCOLO to generate triggers for each group within classes, rather than broadly for each class broadly. This approach leverages reverse engineering and word discriminative analysis to identify potential triggers more effectively. We evaluated this strategy on 10 clean and 10 backdoored models using RoBERTa on the Jigsaw dataset, achieving a 70\% detection accuracy. However, this method relies on the assumption that attackers can accurately identify sensitive attributes, and the accuracy remains suboptimal, highlighting the need for more precise and efficient detection techniques.
\section{Conclusion}

In this paper, we introduce BadFair, a novel, model-agnostic backdoored fairness attack that integrates three key components: \textit{Target-Group Poisoning}, \textit{Non-target Group Anti-Poisoning}, and \textit{Fairness-aware Trigger Optimization}. These techniques enable the model to maintain accuracy and fairness on clean inputs, while surreptitiously transitioning to discriminatory behaviors for specific groups under tainted inputs. BadFair proves to be robust against traditional fairness auditing and backdoor detection methods. On average, it achieves an 88.7\% ASR for the target group, with only a 1.2\% reduction in accuracy across all tested tasks. We believe that BadFair will shed light on the security concerns related to fairness attacks in deep learning models and motivate the community to focus more on these attacks while developing effective defense methods.
\section{Limitations}
The limitations of our paper are as follows: our BadFair is evaluated on popular benchmark datasets and models, including Jigsaw, Twitter, and AgNews datasets; RoBERTa, DeBERTa, and XLNet. However, the paper primarily focuses on classification tasks, potentially constraining the generalizability of our findings to a broader range of NLP tasks such as generation~\cite{chen2023backdoor, xue2024badrag}. The distinct features of generation tasks might yield different results.

\section{Ethical Considerations}
Our findings highlight significant security vulnerabilities in deploying NLP models across critical sectors such as healthcare, finance, and other high-stakes areas. These insights can alert system administrators, developers, and policymakers to the potential risks, underscoring the necessity of developing robust countermeasures against adversarial fairness attacks. Understanding the capabilities of BadFair could spur the development of advanced defense mechanisms, enhancing the safety and robustness of AI technologies. Additionally, a potential defense method is discussed in Section~\ref{sec:defense} to further research into secure NLP application deployment.

\bibliography{custom}

\newpage
\appendix

\twocolumn[\newpage]
\appendix

\noindent\textbf{Target Group and Target Class.} For datasets Jigsaw and Twitter-EEC have been annotated with sensitive attributes for each sentence, while for AgNews, we annotated each sentence by keywords related to \textit{Asia} as belows:


\noindent\texttt{\small [China, India, Japan, South Korea, North Korea, Thailand, Vietnam, Philippines, Malaysia, Indonesia, Singapore, Myanmar, Pakistan, Bangladesh, Sri Lanka, Nepal, Bhutan, Maldives, Afghanistan, Mongolia, Kazakhstan, Uzbekistan, Turkmenistan, Kyrgyzstan, Tajikistan, Saudi Arabia, Iran, Iraq, Israel, Jordan, Lebanon, Syria, Turkey, United Arab Emirates, Qatar, Bahrain, Oman, Kuwait, Yemen, Cambodia, Laos, Brunei, Xi Jinping, Narendra Modi, Shinzo Abe, Lee Hsien Loong, Mahathir Mohamad, Kim Jong-un, Aung San Suu Kyi, Imran Khan, Sheikh Hasina, Salman bin Abdulaziz, Hassan Rouhani, Benjamin Netanyahu, Recep Tayyip Erdoğan, Bashar al-Assad, Genghis Khan, Mao Zedong, Mahatma Gandhi, Dalai Lama, Ho Chi Minh, Pol Pot, King Rama IX, Emperor Akihito, Silk Road, Great Wall, Taj Mahal, Mount Everest, Angkor Wat, Forbidden City, Red Square, Meiji Restoration, Opium Wars, Korean War, Vietnam War, Hiroshima, Nagasaki, Tiananmen, Cultural Revolution, Boxer Rebellion, Gulf War, Arab Spring, ISIS, Persian Gulf, Yellow River, Ganges, Yangtze, Mekong, Himalayas, Kyoto Protocol, Asian Games, Belt and Road, ASEAN, SCO, APEC, SAARC, East Asia Summit, G20 Summit, One Child Policy, Demilitarized Zone]}

\noindent\textbf{Experiment Setting.}
Training times for \BaFair{}, using RoBERTa, varied by dataset: approximately $2$ hour for Jigsaw, $0.4$ hours for Twitter-ECC, and $0.9$ hours for AgNews. For the hyperparameter in our loss function (Equation~\ref{e:1}), we set $\lambda$ to \(|\mathcal{L}_1/\mathcal{L}_2|\) to dynamically maintain the balance.

\noindent\textbf{Fairness Evaluation Metrics.} \label{app:fairness_metrics}
Let \textit{$x_i, y_i, z_i$} as the original input samples, label, and bias-sensitive attribute for every sample \textit{i} in the dataset. \textit{$S(x_i)$} can be represented as sketch sample and \textit{$M(S(x_i))$} is the predicted label $\hat{y_i}$. The true positive rate (\textit{TPR}) and false positive rate (\textit{FPR}) are:
\begin{equation}
    TPR_z = P(\hat{y_i}=y_i|z_i=z)
\end{equation}
\begin{equation}
    FPR_z = P(\hat{y_i}\neq y_i|z_i=z)
\end{equation}

Based on~\cite{li2021estimating, wang2022fairness}, \textit{Statistical Parity Difference (SPD)}, \textit{Equal Opportunity Difference (EOD)}, and \textit{Average Odds Difference (AOD)} are applied to measure and evaluate the fairness. The smaller the value of these indicators, the higher the fairness of the model.

\begin{itemize}

\item \textit{Statistical Parity Difference (SPD)} measures the difference of probability in positive predicted label ($\hat{y}=1$) between protected ($z=1$) and unprotected ($z=0$) attribute groups.
\begin{equation}
    SPD = |P(\hat{y}=1|z=1) - P(\hat{y}=1|z=0)|
\end{equation}

\item \textit{Equal Opportunity Difference (EOD)} measures the difference of probability in positive predicted label ($\hat{y}=1$) between protected ($z=1$) and unprotected ($z=0$) attribute groups given positive target labels ($y=1$). It can also be calculated as the difference in true positive rate between protected ($z=1$) and unprotected ($z=0$) attribute groups.
\begin{equation}
\begin{split}
    EOD &= |TPR_{z=1} - TPR_{z=0}| \\
    &= |P(\hat{y}=1|y=1, z=1) \\
    & - P(\hat{y}=1|y=1, z=0)|
    \end{split}
\end{equation}

\end{itemize}

\noindent\textbf{Evasiveness against Bias Estimation.} 
We investigate the effectiveness of \BaFair{} in evading bias estimation methods and compare with against prior fairness attack SBPA~\cite{jagielski2021subpopulation}. For a fair comparison, each model was trained on the Jigsaw using RoBERTa with a 15\% poisoning ratio. Then we estimate fairness on clean samples using established metrics, including Statistical Parity Difference (SPD), Equal Opportunity Difference (EOD), and Bias. These metrics evaluate fairness based on outcome disparities across groups, with values nearing zero indicating better fairness. The calculations of SPD and EOD are elaborated in Appendix~\ref{app:fairness_metrics}.

\begin{table}[ht!]
\centering
\scriptsize
\setlength{\tabcolsep}{6pt}
\caption{Evaluation of evasiveness against fairness estimation. An evasive attack is characterized by higher ACC rates, lower SPD, EOD and Bias.}
\begin{tabular}{lcccc}\toprule
Attacks & ACC(\%) \(\uparrow\) & SPD(\%) \(\downarrow\) & EOD(\%) \(\downarrow\) & Bias(\%) \(\downarrow\) \\\midrule
Clean Model & 89.3 & 14.3 & 7.43 & 2.67 \\
SBPA & 71.2 & 35.2 & 57.9 & 75.8  \\
\textbf{\BaFair{}} & \textbf{88.4} & \textbf{18.5} & \textbf{8.21} & \textbf{3.15} \\\bottomrule
\end{tabular}
\label{t:fairness_defense}
\end{table}

The results in Table~\ref{t:fairness_defense} show that all the fairness metrics are similar between \BaFair{} and clean models. The underlying reason is that the fairness attack in \BaFair{} is only activated by the trigger, so the fairness audition cannot detect such attack on clean dataset. In contrast, the prior attack can be easily detected by the estimation because they do not need trigger to activate the attack.

\noindent\textbf{Ablation Study on Poisoning Ratio \bm{$\gamma$}}. The poison ratio defines the percentage of data associated with an attached trigger, which impacts the performance of \BaFair{}. To demonstrate the impact, we evaluated \BaFair{} across a range of poisoning ratios, from 1\% to 30\%, as shown in Table~\ref{t:poison_ratio}. Remarkably, even with a minimal poisoning ratio of 1\%, \BaFair{} achieves a substantial PBias score of 22.6\%, while obtaining a high T-ASR of 82.2\%. Particularly, when $\gamma$ is set to 15\%, \BaFair{} achieves an impressive T-ASR of 91.1\% with a mere 0.9\% CACC loss. Furthermore, \BaFair{} consistently maintains a high clean accuracy across all tested poisoning ratios.

\begin{table}[ht!]
\centering
\scriptsize
\setlength{\tabcolsep}{2.5pt}
\caption{\TrojFair{} performance across various poisoned data ratios.}
\begin{tabular}{lccccccc}\toprule
\multirow{2}{*}{\makecell[c]{Poisoning\\Ratio (\%)}} & \multicolumn{2}{c}{Clean Model} & \multicolumn{5}{c}{Poison Model} \\\cmidrule(lr){2-3}\cmidrule(lr){4-8}
 & ACC & Bias & CACC$\uparrow$ & CBias$\downarrow$ & T-ASR$\uparrow$ & NT-ASR$\downarrow$ & PBias$\uparrow$ \\\midrule
1 & 89.3 & 2.67 & 89.1 & 2.70 & 82.2 & 42.3 & 22.6 \\
5 & 89.3 & 2.67 & 88.9 & 2.81 & 84.9 & 27.3 & 49.4 \\
15 & 89.3 & 2.67 & 88.4 & 3.15 & 91.1 & 21.8 & 45.5 \\
30 & 89.3 & 2.67 & 87.6 & 3.32 & 93.2 & 13.5 & 59.8 \\\bottomrule
\end{tabular}
\label{t:poison_ratio}
\end{table}

\noindent\textbf{Datasets.} Details of the datasets, such as classification tasks, number of classes, training sample sizes, and test sample sizes are presented in Table~\ref{t:dataset}.
\begin{table}[h!]
\centering
\scriptsize
\setlength{\tabcolsep}{4pt}
\caption{Dataset Characteristics.}
\begin{tabular}{llccc}\toprule
Dataset & Task & Classes & Train-set & Test-set \\\midrule
Jigsaw & Toxicity detection & 2 & 180,487 & 9,732 \\
Twitter-EEC & Sentiment Classification & 2 & 6,000 & 2,000 \\
AgNews & News Topic Classification & 4 & 120,000 & 7,600\\\bottomrule
\end{tabular}
\label{t:dataset}
\end{table}




\end{document}